\documentclass{aa}  

\usepackage{graphicx}
\usepackage{multirow}

\usepackage{txfonts}
\usepackage{natbib,twoopt}

\begin{document}

   \title{Uncertainties in the solar photospheric oxygen abundance}

   \author{M. Cubas Armas
          \inst{1,2}
          \and
          A. Asensio Ramos\inst{1,2}
          \and
	  H. Socas-Navarro\inst{1,2}
	  }	
   \institute{Instituto de Astrof\'isica de Canarias (IAC), Avda V\'ia L\'actea S/N,
              38200 La Laguna, Tenerife, Spain\\
              \email{mcubas@iac.es}
         \and
             Departamento de Astrof\'isica, Universidad de La Laguna, 38205 La Laguna, Tenerife, Spain\\
             }

   \date{Received ; accepted }

 
  \abstract
  {} 
   {The purpose of this work is to better understand the confidence limits of the photospheric solar oxygen abundance derived from three-dimensional models using the forbidden [OI] line at 6300 \AA , including correlations with other parameters involved.}
   {We worked with a three-dimensional empirical model and two solar intensity atlases. We employed Bayesian inference as a tool to determine the most probable value for the solar oxygen abundance given the model chosen. We considered a number of error sources, such as uncertainties in the continuum derivation, in the wavelength calibration and in the abundance/strength of Ni.}
   {Our results shows correlations between the effects of several parameters employed in the derivation. The Bayesian analysis provides robust confidence limits taking into account all of these factors in a rigorous manner. We obtain that, given the empirical three-dimensional model and the atlas observations employed here, the most probable value for the solar oxygen abundance is $\log(\epsilon_O) = 8.86\pm0.04$. However, we note that this uncertainty does not consider possible sources of systematic errors due to the model choice.}
   {}


   \maketitle
%

\section{Introduction}

   The solar chemical composition is still under debate, being particularly important the case of the oxygen (the third most abundant element in the Sun after hydrogen and helium). One of the abundance set traditionally accepted considers an oxygen abundances of log($\epsilon_O$) = 8.93$\pm$0.04 \citep{1984A&A...141...10G}, in the astronomical usual scale referred to hydrogen (where $\log(\epsilon_O)  = log(N_X/N_H)+12$). This abundance was later downward revised to log($\epsilon_O$) = 8.83$\pm$0.06 \citep{1998SSRv...85..161G}. With this abundance set, there is excellent agreement between solar interior models predictions and the helioseismology. However, using more recent three-dimensional theoretical models, lower solar metallicities have been obtained. For example, \cite{2004A&A...417..751A} reached log($\epsilon_o$) = 8.66$\pm$0.05, which may spoil the agreement with the helioseismology \citep[see][]{2008PhR...457..217B}. Note however that a recent study suggests that we might be underestimating interior opacities which could restore the agreement with helioseismology \citep{bailey}.  Given that oxygen is a very important element for stellar interior models, adopting such a dramatic revisions would have implications over a broad range of topics in astrophysics. 

Further works have not yet converged to a satisfactory resolution of the issue. On one hand, some  studies found a relatively low oxygen abundance \citep[e.g.,][]{2007ApJ...660L.153S,2009A&A...508.1403P,2010Ap&SS.328..179G}; but, on the other hand, high oxygen abundance has also been obtained \citep[e.g.,][]{2008ApJ...682L..61C,2008ApJ...686..731A,2015A&A...577A..25S}. Perhaps an intermediate value, such as log($\epsilon_O$) $\approx$ 8.75 \citep[e.g.,][]{2008A&A...488.1031C,2015A&A...579A..88C} might be able to satisfy all of the constraints if the uncertainties had been underestimated thus far (i.e., the model atmosphere is typically assumed to be perfect except in Socas-Navarro 2011). In conclusion, the oxygen abundance issue remains open and largely controversial.

A key factor in the abundance derivation is the solar atmosphere model used. It seems natural that a three-dimensional model should be preferred over a one-dimensional one. However, when the three-dimensional model is a numerical simulation and the one-dimensional model has been obtained empirically by fitting observations, it is not clear that the former is better than the latter to fit observations  \citep{2008ApJ...686..731A}. 

Deriving element abundances is far from trivial. The results are not determined solely by the solar atmosphere model used. They are also dependent of the solar atlas adopted to fit the model and even some details on the calibration (choice of the continuum level, spectrum rectification, wavelength calibration, etc). Recently, some studies had noted the differences between the solar atlases  \citep[e.g.,][]{2008A&A...488.1031C,2009A&A...498..877C,2016arXiv160403748D} and the need to use more than one in the analysis. Another obvious factor is the value of the line strength employed (parameterized in terms of $\log(gf)$), which have been discussed in some works \citep[e.g.,][]{2000MNRAS.312..813S,2003ApJ...584L.107J}.

In this paper, we present a study of the oxygen abundance in the solar photosphere with the novelty that we applied Bayesian inference to properly disentangle the effects of the relevant parameters involved. In order to do that, we used a three-dimensional empirical model of the solar atmosphere \citep{2011A&A...529A..37S,2015A&A...577A..25S} to fit the forbidden [O~{\sc i}] line at 630.03 nm
observed in two different solar atlases. In Section 2, we describe the solar model used, the syntheses made and the solar atlas observations. A brief summary of the Bayesian inference and a table with our priors are given in Section 3. Then, we present the results in Section 4 and, finally, some conclusions in Section 5.
\section{Model atmosphere and observational data}
\subsection{Solar model used and synthesis}
The solar atmosphere model that is used in this work was derived  
by \cite{2011A&A...529A..37S,2015ascl.soft08002S} based on observations
of the spectro-polarimeter \citep[SP;][]{lites_hinode01} of the solar optical telescope onboard 
the Hinode satellite \citep{kosugi_hinode07}. The observations have a field of view very close 
to the disk center and the wavelength range covers from 6300.89 to 6303.27~\AA. The spectrum 
at each pixel was inverted using the code NICOLE \citep{2015A&A...577A...7S} 
to determine a column with the height stratification of temperature, line-of-sight velocity and magnetic field vector. The reader is referred to the papers cited above for more details on the model.

Our aim is to compare the synthetic line profiles with those observed in 
different atlases of the solar spectrum. The spatial resolution of these
observations is extremely poor (of the order of tens of arcsec). Thus, we
mimick this resolution by averaging all profiles synthesized in all
pixels of the snapshot. As a consequence, the forward model in this three-dimensional snapshot turns out to be
very time consuming. Given that the Bayesian analysis that we explain in the following
section requires the evaluation of several tens of thousand of forward models, it is convenient to pre-compute a grid
of models and carry out the Bayesian inference using a simulator that
just interpolates on the grid of models \citep{ohagan06}.
This greatly accelerates the inference by several orders of magnitude.

The precomputed database is built using a cartesian grid in all parameters of the
model. The model parameters can be separated in two different classes: slow parameters,
that require the recomputation of the synthesis in the whole snapshot because they
affect the radiative transfer, and fast parameters, that can be applied directly to the
profiles with simple manipulations. The first two slow parameters are 
the oxygen and nickel abundances 
(note that the [O \textsc{i}] feature analyzed here has a Ni \textsc{i} blend). The third
slow parameter is an enhancement factor for the velocities in the lower layers of the
snapshot. This parameter is introduced because the model of \cite{2011A&A...529A..37S}
is constructed fitting the strong Fe \textsc{i} lines at 630.1 and 630.2 nm, and it misses some of 
the dynamics in the lower layers of the atmosphere, precisely where the oxygen line forms. 
Fortunately, the nearby Sc \textsc{ii} line, very similar in strength and formation height to the [O \textsc{i}] line, 
is an excellent calibration tool for this missing turbulence, which we parameterize as a multiplicative 
enhancement factor applied on the velocities of the lower layers. For more details, see the 
discussion in \cite{2011A&A...529A..37S}. The fast parameters are: two for a linear correction
to the continuum, a global velocity shift that takes into account imprecision in the wavelength
calibration (this correction is important because a degeneration exists between
the ratio of Ni and O abundances,this is explained in more detailed in sect 4.2 below) and the uncertainty of the fit.

To end up with a well sampled database, we took ten possible values for each one of the slow
parameters, resulting a grid of 1000 models. The oxygen and nickel 
abundances were varied between typical values reported in the literature \citep[see e.g.,][]{1984A&A...141...10G,1989GeCoA..53..197A,1998SSRv...85..161G,2009ARA&A..47..481A}. 
Their minimum and maximum values are shown in Table \ref{table:priors} together with 
those for the enhancement factor.

It is important to stress that we considered the nickel abundance in the analysis because 
the spectral feature at 630.0 nm is actually the result of a blend between the [O~{\sc i}] line at
630.03 nm and a Ni~{\sc i} line at 630.03 nm. We considered the two major isotopes $^{58}$Ni~{\sc i} and $^{60}$Ni~{\sc i}, with a log(gf) value of -2.11 and a fraction of 72\% and 28\% respectively for each isotope (\cite{2003ApJ...584L.107J}; see Table \ref{table:atomic}). 

For the synthesis with NICOLE , we chose a wavelength range from 630.0 to 630.3 nm, 
including also the Fe~{\sc i} lines at 630.15 and 630.25 nm. We introduced the Fe~{\sc i} lines 
because the [O~{\sc i}] line is formed in the far wing of the 630.15 nm Fe~{\sc i} line 
and we need to include this effect in our synthesis, since it affects the continuum
estimate. 

We introduce an additional simplification in the forward modeling, given that the model of 
\cite{2011A&A...529A..37S} has $200 \times 200$ spatial pixels in the field of view. Synthesizing the wavelength range in each one of the 1000 models in our grid 
would be very demanding computationally.
For this reason, we carried out a Montecarlo selection of a subset of 1000 pixels, as a good compromise between accuracy and computational time, that
produce an average profile that is ``equivalent'' to the average profile of the whole
snapshot. By ``equivalent'' we mean that the differences between the average intensity profile of 
the full map and that of the subset is smaller than $0.10\%$ in the oxygen line.

In summary, we constructed a grid of 1000 models where each model has 1000 pixels. For each one of 
them, we synthesized the spectral lines of interest using NICOLE, taking into account the atomic information compiled 
in Table ~\ref{table:atomic}. For the sake of reproducibility, the table also includes the references for 
the log($gf$) values. The log(gf) value for the Fe~{\sc i} at 630.25 nm was taken from 
\cite{2011A&A...529A..37S}. The radiative, Stark and van der Waals damping parameters are also provided 
for each line except the Fe lines. For those, we give instead the $\alpha$ and $\sigma$ damping 
parameters using the method of \cite{1995MNRAS.276..859A} obtained with the code of \cite{1998PASA...15..336B}.

\begin{table*}
\centering                          
\begin{tabular}{c c c c c c c c c c c}        
\hline\hline                 
Ion &  Wavelength [nm] & Ex. Pot. [eV] & log(gf) & Conf. (lower) & Conf. (Upper) & $\gamma_{rad}$  & $\gamma_{Stark}$ & $\gamma_{Waals}$ & $\sigma$ & $\alpha$\\
\hline  
[O~{\sc i}]  & 630.0304  & 0.000 & -9.717$^a$ & $^3P_2$ & $^1D_2$ & 0.0 & 0.05  & 1.00 & ... & ...\\
Ni~{\sc i}   & 630.0335  & 4.266 & -2.253$^b$ & $^3D_1$ & $^3P_0$ & 2.63& 0.054& 1.82 & ... & ...\\
Ni~{\sc i}   & 630.0355  & 4.266 & -2.663$^b$ & $^3D_1$ & $^3P_0$ & 2.63& 0.054& 1.82 & ... & ...\\
Sc~{\sc ii} & 630.0678   & 1.507 & -1.898$^c$ & $^3P_2$ & $^3D_2$ & 2.30& 0.05 & 1.30 & ... & ...\\
Fe~{\sc i}  & 630.15012 & 3.654 & -0.718$^c$ & $^5P_2$ & $^5D_2$ & ... & ... & ... & 834.4 & 0.243\\
Fe~{\sc i}  & 630.24940 & 3.686 & -1.13   & $^5P_1$  & $^5D_0$ & ... & ... & ... & 850.2 & 0.239\\
 \hline                                   
\end{tabular}
\caption{Adopted atomic parameters. $\gamma_{rad}$, $\gamma_{Stark}$ and $\gamma_{Waals}$ are the radiative, the Stark and the van der Waals damping parameters (units $10^8$ $rad$ $s^{-1}$). The log($gf$) are taken from: $^a$ \cite{2000MNRAS.312..813S}, $^b$ \cite{2003ApJ...584L.107J}, $^c$ VALD database.}   
\label{table:atomic}      
\end{table*}

\subsection{Observational data}
In this study we also intend to assess the influence of the choice of solar observation on the abundance determination. 
We used two different intensity atlases of the solar photosphere which have a good 
signal-to-noise (S/N) ratio and have been widely used in previous works 
\citep[e.g.][]{2008A&A...488.1031C, 2009A&A...498..877C, 2008ApJ...686..731A, 2009PASA...26..345M}.  

The first one is the disk center intensity atlas of the solar spectrum from 3000~{\AA} to 10000~{\AA} by 
\cite{1973apds.book.....D}. The spectrum was obtained at the International Scientific Station of the Jungfraujoch 
and may be downloaded from the BASS2000 web server\footnote{http://bass2000.obspm.fr/solar\_spect.php}.
The second atlas that we considered was the Kitt Peak intensity FTS (Fourier Transform Spectrometer) atlas by
\cite{1984SoPh...90..205N}. This atlas was produced with the FTS instrument at the McMath telescope and 
spans the wavelength range from 3290 to 12510~{\AA}.


\section{Bayesian analysis}
We employ the formalism of Bayesian inference \citep[e.g.,][]{gregory05} to quantify the uncertainty in our conclusions derived from the use of a prescribed model, a bunch of free parameters and a set of priors (information known a priori). The aim of the
Bayesian analysis is to compute the posterior probability distribution associated with all model parameters by taking
into account the information provided by the observations and all our a-priori information. It relies on two fundamental
tools. The first one is the Bayes theorem, that describes very simply how the prior information is updated with the
acquisition of new observations, to the posterior information.
If we denote the model parameters of interest with the vector $\boldsymbol{\theta}$, and the observations with $D$, the
Bayes theorem states that
\begin{equation}
p(\boldsymbol{\theta}|D,I) = \frac{p(\boldsymbol{\theta}|I)p(D|\boldsymbol{\theta},I)}{p(D|I)}.
\label{eq:bayes1}
\end{equation}
In the Bayes theorem, $p(\boldsymbol{\theta}|I)$ is the prior probability distribution of $\theta$, that encodes 
all the a-priori information we know about the parameters (i.e., if a quantity is positive, that a certain region of
the space of parameters is more probable than others, etc.). $p(D|\boldsymbol{\theta},I)$ is the likelihood, which
encodes the information about the model parameters that can be extracted from the observations. $p(\boldsymbol{\theta}|D,I)$
is the posterior distribution, that describes everything that we know about the model parameters. $p(D|I)$ is
the model evidence, which is an unimportant constant in our case, because it does not depend on $\boldsymbol{\theta}$.
Finally, $I$ refers to any important context information that is necessary for the inference. In our case, for instance,
the specific radiative transfer model that we use. The second fundamental tool is the
marginalization, that is used to obtain the posterior distribution for any parameter
taking into account the uncertainties and correlations with other parameters:
\begin{equation}
p(\theta_1|D,I) = \int \mathrm{d}\theta_2 \ldots \mathrm{d}\theta_{n} p(\boldsymbol{\theta}|D,I)
\end{equation}

In this work we considered seven free parameters, that we consider necessary for
explaining all the expected variability of the line profiles: the oxygen abundance (which is the main parameter of interest), the abundance of nickel (necessary to reproduce the line shape because the oxygen line of interest is blended with
a Ni \textsc{i} line), a global wavelength shift to account for possible inaccuracies in the wavelength scale, an enhancement
factor for the dynamics that is described above, two free parameters that are used to correct the continuum level 
and its slope, and finally a parameter $\sigma$ that describes the (unknown) uncertainty of the observations and also
quantifies the quality of our modeling.

In any Bayesian inference, it is crucial to define the priors. In our case, we define the priors as flat (meaning there is no preference) for almost all parameters in the ranges shown in Table~\ref{table:priors}. The ranges have been chosen taking into consideration the results obtained in previous studies \citep[see e.g.,][]{1984A&A...141...10G,1989GeCoA..53..197A,1998SSRv...85..161G,2009ARA&A..47..481A} but 
trying not to discard parts of the space of parameters that might be compatible with
the observations. The only non-flat prior is that of $\sigma$, for which we
adopted a Jeffreys prior, $p(\sigma) \propto \sigma^{-1}$, because it is a scale parameter
that can potentially have values spanning several orders of magnitude \citep[e.g.,][]{gregory05}. 

For computing the likelihood, we assume the following generative model:
\begin{equation}
O(\lambda_i) = I(\boldsymbol{\theta},\lambda_i) + n(\lambda_i),
\end{equation}
which states that the observations for the $i$-th wavelength point, $O(\lambda_i)$, 
can be modeled with $I(\boldsymbol{\theta},\lambda_i)$ plus some
uncertainty that has zero mean and variance $\sigma^2$. Following the standard
approach \citep[e.g.,][]{gregory05}, the likelihood has the form of an uncorrelated
multivariate normal distribution, which reduces to:
\begin{equation}
p(D|\boldsymbol{\theta},I) = \prod_{i=1}^n \frac{1}{\sqrt{2\pi}\sigma} \exp\left[ 
-\frac{-\left[O(\lambda_i) - I(\boldsymbol{\theta},\lambda_i)\right]^2}{2\sigma^2} \right],
\label{eq:bayes2}
\end{equation}
where $n$ is the number of observed wavelengths.

We sample the posterior distribution using the nested sampling tecnique by \cite{2004AIPC..735..395S}
\footnote{We use the implementation of the Python package \texttt{nestle}, which
can be obtained from \texttt{http://kbarbary.github.io/nestle/}.}.

\begin{table}
\centering                          
\begin{tabular}{lll}        
\hline\hline                 
Parameter &  Range & Type\\
\hline  
log($\epsilon_o$)  & (8.55, 9.20) & Flat\\
log($\epsilon_{Ni}$)  & (5.84, 6.36) & Flat\\ 
v [km/s] & (-1, 1) & Flat\\
Enhancement factor & (0.5, 2.3) & Flat\\
$\sigma$ & (0.0001, 0.02) & Jeffreys\\
Cont. slope [$I/I_c$ $\dot{}$ $\AA{}^{-1}$] & (-0.1, 0.1) & Flat\\
Cont. y intercept [$I/I_c$] & (0.9, 1.1)  & Flat \\    
 \hline                                   
\end{tabular}
\caption{Priors selected for each parameter of our analysis.}    
\label{table:priors}      
\end{table}

\section{Results}

\subsection{Line fit}
   
\begin{figure*}     
  \includegraphics[width=.5\hsize]{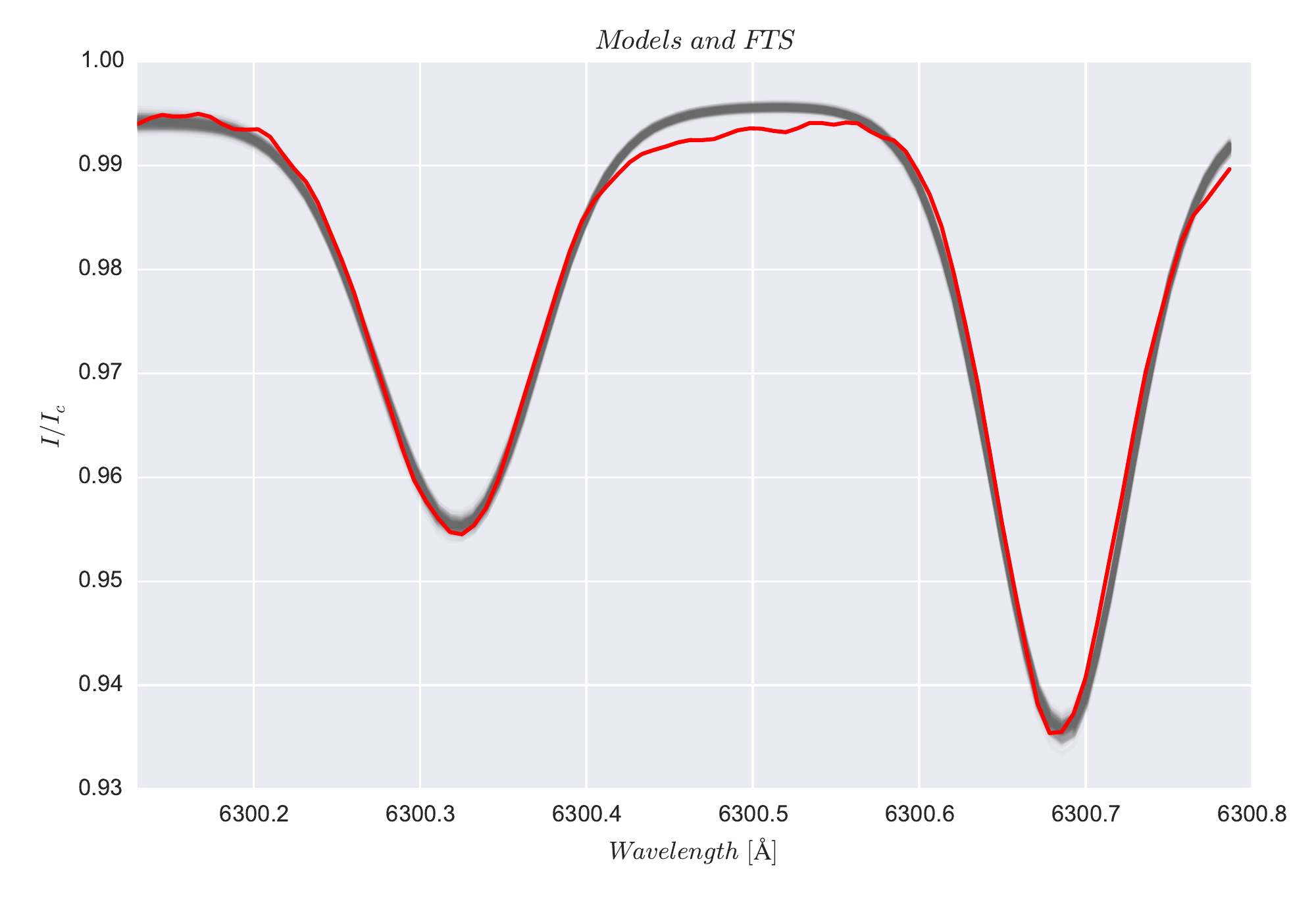}  \includegraphics[width=.5\hsize]{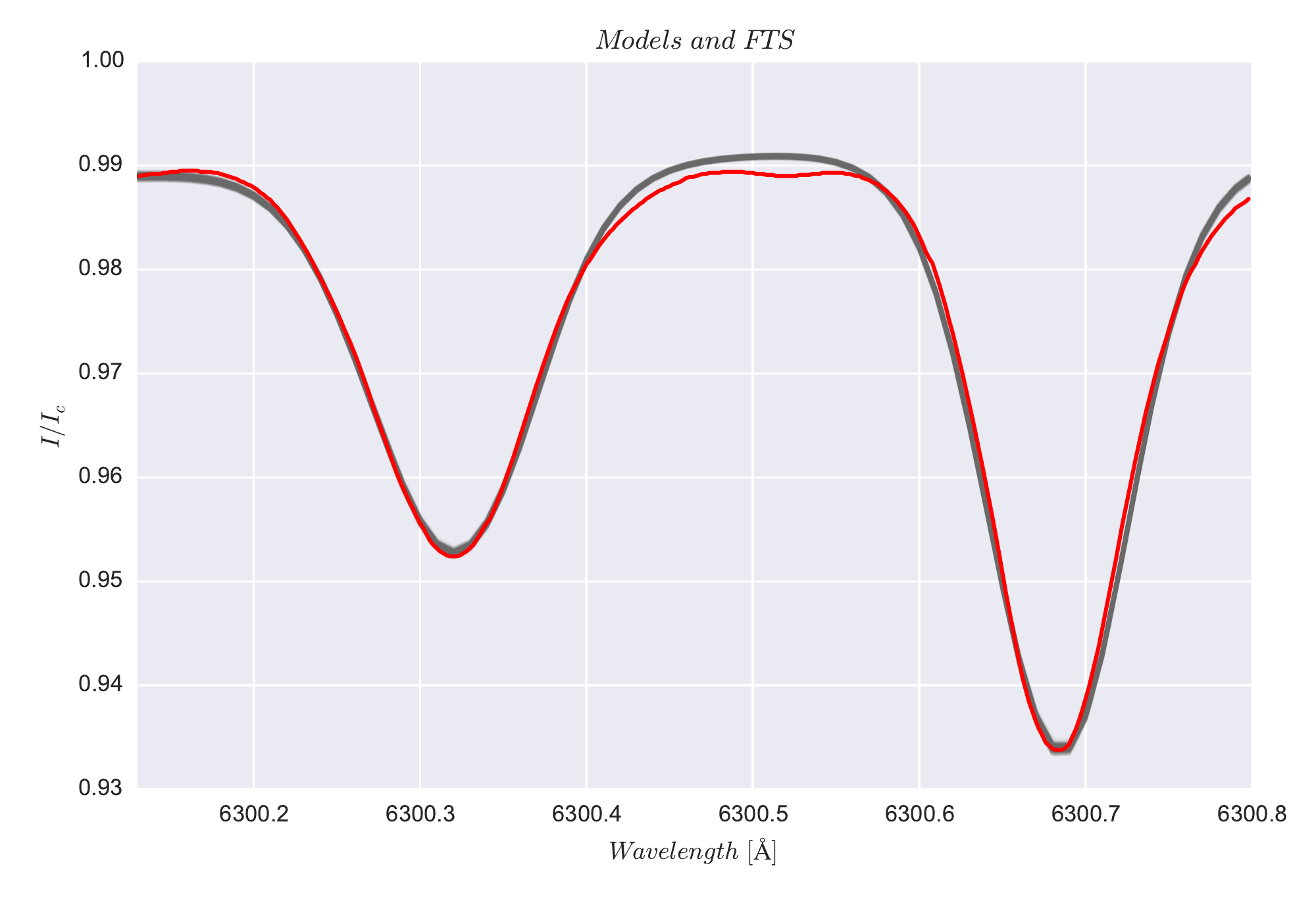} 
  \caption{Some representative fits of the [O~{\sc i}] line (left) and Sc~{\sc ii} line (right) are shown in black. The blue profile correspond to the Neckel solar atlas in the left panel and to the Delbouille atlas in the right panel.}
   \label{fig:prof_neck_pior}
   \end{figure*}

Figure~\ref{fig:prof_neck_pior} shows our fits of the [O~{\sc i}] and Sc~{\sc ii} lines in our 
model (black lines) with respect to the atlases (blue). The left panel uses the   
Neckel atlas and the right one uses the Delbouille atlas. Instead of a single fit, we provide
a sampling of models that are compatible with the observations. They are obtained by synthesizing
models with parameters extracted from the posterior distribution.
Figure \ref{fig:prof_neck_pior} clearly shows that our model fits very well both atlases, even in very weak 
lines like these. However, the models present some discrepancies with the observations in the continuum region in
between the two lines, where the models set the continuum to a higher level. Given that we have this effect in both atlases, 
this might be produced by some unknown absorption in this region. The lines
of interest are very weak, thus, this region could affect our abundance determination indirectly by setting the continuum level
on a wrong level. We tried to minimize this effect by introducing the Sc~{\sc ii} line in the analysis, which
then sets a much better continuum level estimation. We have plans to refine this approach in the
future by using a modeling that is able to absorb these deficiencies 
\citep[similar to the nonparametric approach based on Gaussian Processes of][]{2015ApJ...812..128C}.

\subsection{Marginal posterior distributions}
\begin{figure*}            
  \includegraphics[trim={4cm 1cm 1cm 3cm},clip,width=1.10\hsize]{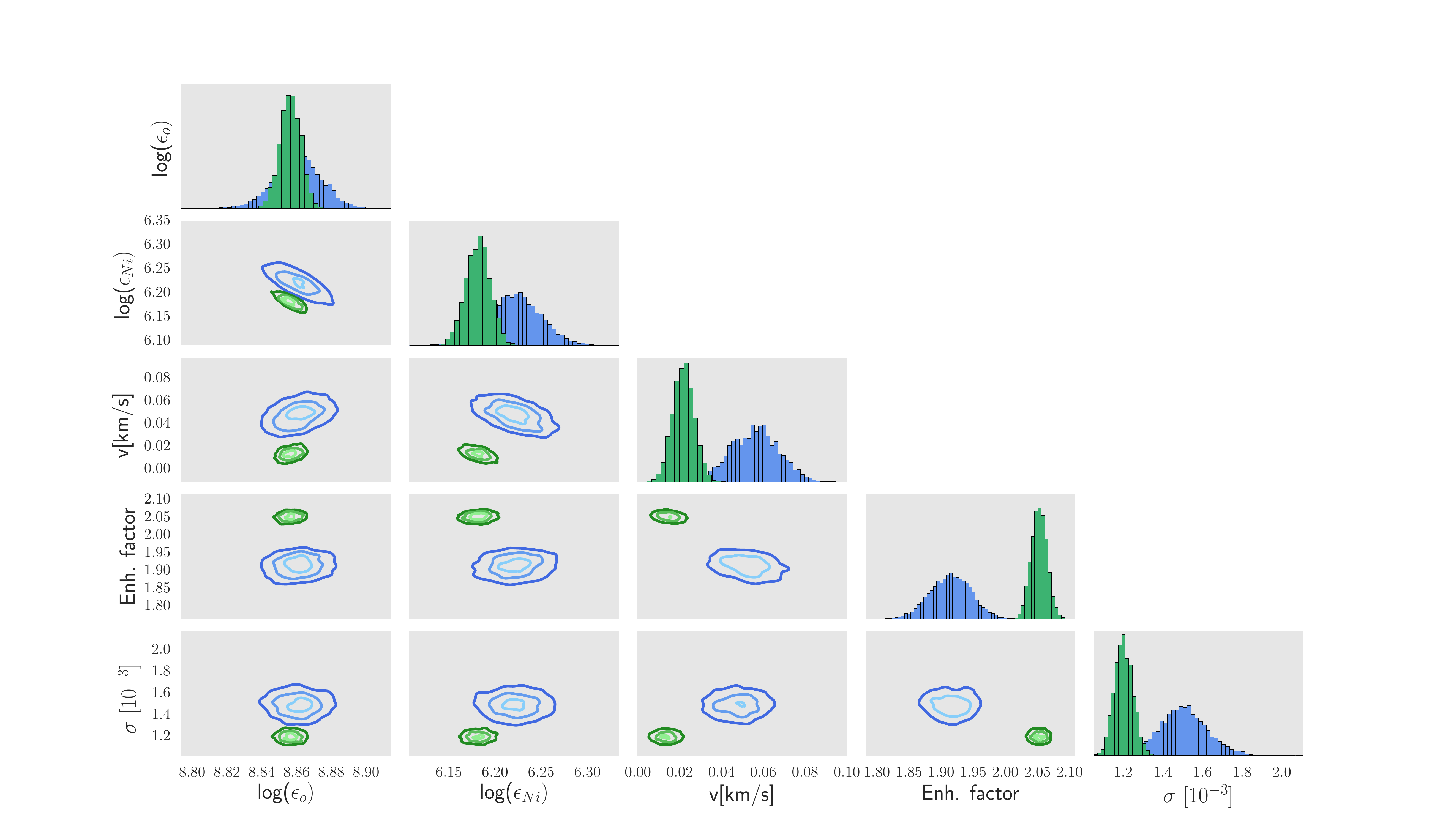}   
  \caption{Histograms and correlations of the posterior distributions of the main five parameters (in blue for Neckel solar atlas and green for Delbouille atlas). From left to right, and top to bottom, we show: oxygen abundance, nickel abundance, velocity (in km/s), enhancement factor and $\sigma$.}
   \label{fig:corner}
   \end{figure*}

The joint and marginal  posterior distributions for the five main parameters (oxygen and nickel abundances, 
velocity, enhancement factor and $\sigma$) are shown in Fig.~\ref{fig:corner}. The blue distributions 
correspond to the Neckel atlas while the green distributions correspond to the Delbouille atlas. We did not show 
the distribution for the two parameters of the continuum because they are nuisance parameters that are only
needed to have a good fit of the line, but they do not provide relevant information.

For both atlases, the joint and marginal posterior distributions show a Gaussian-like shape, where those corresponding 
to the Neckel atlas always present a larger width. This width indicates that the uncertainties in the abundances
inferred from the Neckel atlas are larger than those obtained from the Delbouille atlas. Moreover, the marginal
distribution for $\sigma$ also shows that this parameter is smaller for the case of the Delbouille atlas. Thus, our model
seems to provide more constrained model parameters and better fits for the Delbouille atlas.

It is encouraging that the marginal posterior for the oxygen abundance is consistent in both atlases, with that
inferred from the Delbouille atlas being contained inside the uncertainty associated with the Neckel atlas. 
It is interesting that the maximum marginal a posteriori (the peak of the marginal posterior)
is located at roughly the same value for the two cases. 

This is not the same with the rest of parameters. The marginal posteriors corresponding to the Delbouille atlas tend 
to be shifted towards smaller values, except for the enhancement factor, where the opposite happens. In this last
parameter, the marginal distributions even indicate that the inferred enhancement factors are
not compatible (at least to 3 standard deviations).

If we look at the panels with the joint posteriors (showing the correlation 
between pairs of parameters), we see a clear correlation between the abundances of oxygen and 
nickel. This correlation is a direct consequence of the fact that the [O \textsc{i}] line
is a blend: we can obtain a fit equally good if we take a model with a higher nickel abundance 
and a lower oxygen abundance and viceversa. Furthermore, these two parameters (oxygen and nickel abundances) 
are correlated with the velocity shift. This correlation makes sense for small velocities, since a wavelength
shift displaces the line and one can still obtain a good fit by appropriately modifying the abundances. For example, 
a blue shift can be compensated with a lower nickel abundance and a higher oxygen abundance. 

All marginal posterior distributions have a Gaussian-like shape. Therefore, we summarize them in 
Table \ref{table:median} by providing the median and the uncertainty defined by the percentiles 16 and 84 (equivalent
to the standard $1\sigma$ uncertainty in the Gaussian case). It is clear in the table that the most probable 
value for the oxygen abundance is the same (down to the second decimal) in both atlases. The value obtained from 
this study is, hence, $\log(\epsilon_O) = 8.86\pm0.01$. This value is within the category of high solar oxygen 
abundances: lower than the $8.93\pm0.02$ value in \cite{1984A&A...141...10G}, but compatible with 
the $8.83\pm0.06$ revision in \cite{1998SSRv...85..161G}.

\begin{table*}
\centering          
\begin{tabular}{c c c c c c}     
\hline\hline       
Atlas & log($\epsilon_o$) & log($\epsilon_{Ni}$) & v [km/s] & Enh. factor & $\sigma$\\
\hline                    
Neckel & $8.861\pm0.014$ & $6.22\pm0.03$ & $0.05\pm0.01$ & $1.92\pm0.03$ & $0.0015\pm0.0001$\\
Delbouille & $8.856\pm0.006$ & $6.18\pm0.01$ & $0.014\pm0.005$ & $2.05\pm0.01$ & $0.00120\pm0.00005$\\
\hline                  
\end{tabular}
\caption{Parameters median values and deviations for both atlases. }             
\label{table:median}      
\end{table*}

\subsection{More experiments}

In order to quantify how much the impact of the continuum level and the correlation between the oxygen and 
nickel abundances are, we carried out two more studies. They are intended also to check
the robustness of our results.

\begin{table*}
\centering          
\begin{tabular}{c c c | c c c c c}     
\hline\hline       
 \multicolumn{3}{c |}{} &  log($\epsilon_o$)  &  log($\epsilon_{Ni}$)  & v [km/s] & Enh. factor & $\sigma$\\
 \hline                    
  \multirow{4}{*}{Both lines}  & \multirow{2}{*}{Neckel} & prior Ni & $8.864\pm0.013$ & $6.21\pm0.03$ & $0.05\pm0.01$ & $1.92\pm0.03$ & $0.0015\pm0.0001$\\
  & & no prior & $8.861\pm0.014$ & $6.22\pm0.03$ & $0.05\pm0.01$ & $1.92\pm0.03$ & $0.0015\pm0.0001$\\
  \cline{2-8}
  & \multirow{2}{*}{Delbouille} & prior Ni & $8.858\pm0.006$ & $6.18\pm0.01$ & $0.014\pm0.005$ & $2.05\pm0.01$
 & $0.00120\pm0.00005$\\
  & & no prior & $8.856\pm0.006$ & $6.18\pm0.01$ & $0.014\pm0.005$ & $2.05\pm0.01$ & $0.00120\pm0.00005$\\
  \hline
   \hline
 \multirow{4}{*}{Only [O~{\sc i}]} & \multirow{2}{*}{Neckel} & prior Ni & $8.910\substack{+0.015 \\ -0.019}$ & $6.02\pm0.05$ & $0.17\substack{+0.03 \\ -0.04}$ & $1.70\pm0.04$ & $0.00052\pm0.00005$\\

  & & no prior & $8.940\substack{+0.007 \\ -0.012}$ & $5.90\substack{+0.06 \\ -0.04}$ & $0.23\substack{+0.01 \\ -0.02}$ & $1.74\pm0.04$
 & $0.00050\pm0.00005$\\
  \cline{2-8}
  & \multirow{2}{*}{Delbouille} & prior Ni & $8.937\substack{+0.004 \\ -0.007}$ & $5.88\substack{+0.04 \\ -0.03}$ & $0.171\substack{+0.007 \\ -0.013}$ & $1.99\pm0.02$ & $0.00036\pm0.00005$\\
  & & no prior & $8.940\pm0.002$ & $5.85\substack{+0.02 \\ -0.01}$ & $0.178\substack{+0.003 \\ -0.004}$
 & $2.00\pm0.01$ & $0.00036\pm0.00002$
\\
\hline                  
\end{tabular}
\caption{Comparison of the results for the different experiments. }   
\label{table:experiments}                
\end{table*}

First, we carried out the same analysis as before but reducing the wavelength range to just take into 
account the [O~{\sc i}] line. Although we do not display the fits or the marginal posteriors for simplicity, the 
summary of the results is displayed in Table~\ref{table:experiments} (see rows corresponding labeled as "Only [O~{\sc i}]" and 
"no prior"). For completeness, we repeat in this table the results of the previous section for an easier comparison 
(labeled as "Both line" and "no prior"). Thus, modifying the range of wavelengths, the most probable oxygen abundance
changes from log($\epsilon_O$) = 8.86, a value compatible with \cite{1998SSRv...85..161G}, to 
log($\epsilon_O$) = 8.94, a value compatible with \cite{1984A&A...141...10G}. 
Because we are neglecting the wing of the Sc \textsc{ii} line and the inferred abundance
is different, we conclude that a good estimate of the continuum is crucial for the estimation of abundances. 
However, this large oxygen abundance seems improbable at the light of the really low nickel abundance inferred which is
incompatible with all previous reports in the literature (that usually give $\log(\epsilon_\mathrm{Ni}) >6$).

The second test is motivated by the low nickel abundance of the previous results. In this case, we redo the analysis
but including a different prior for the nickel abundance. This prior is set to a Gaussian distribution
with mean $\mu = 6.17$ and standard deviation of $\sigma = 0.07$, as reported in \cite{2009ApJ...691L.119S}. We
intentionally increase the value of $\sigma$ to decrease the informativeness of the prior and let the
data drive the results. The results of this new analysis are also shown in Table~\ref{table:experiments} (with labels "prior Ni"). 
The results show that the inference only with the [O \textsc{i}] line are not very reliable because, when including the 
Gaussian prior for the nickel abundance, the modification in the oxygen and nickel abundances are greater than the 
previously quoted uncertainties. This fundamentally means that the result is strongly
dependent on the prior. However, this is not the case when using the [O \textsc{i}] and Sc \textsc{ii} lines
together, where we obtain results that are essentially insensitive to the prior.

\subsection{Abundance and log($gf$) factor}
We are aware that an important source of uncertainty in the determination
of abundances are the atomic parameters of the spectral lines, specifically
the value of log($gf$). We have tried to use the most recent determinations, but it
is true that a slightly different value of log($gf$) inside its uncertainty would produce
different oxygen and nickel abundances. Therefore, one can consider our determination of the abundance
to be indeed an inference over the product $gf \epsilon_O$, which remains valid for weak lines.

The $gf$ value used in this work \citep{2003ApJ...584L.107J} is accepted as the most accurate, but its uncertainty
can go up to 10\%. This uncertainty translates into $log(gf) = -9.717\pm0.043$. If this is taken into
account, it would induce an uncertainty in the estimated oxygen abundance of $log(\epsilon_O) = 8.86\pm0.04$. 

\section{Conclusions}
There are many parameters involved in the determination of abundances. To end up with an accurate determination 
of abundances, it is crucial to put emphasis on appropriately considering all of them. This is the motivation
of this work and it represents a first step towards a reliable determination of the solar abundances. Several
conclusions can be extracted from our work:

\begin{itemize}
\item We have used a very flexible generative model that contains seven parameters. Some of these
parameters are nuisance parameters that are of no diagnostic interest but which are necessary for 
a good modeling of the spectral line. Including these nuisance parameters and marginalizing over
them is crucial for a reliable determination of abundances. 
\item Including nearby spectral lines turns out to be very important, because they modify the continuum level. We have found that a good characterization of the continuum level in these weak lines is crucial for the inference of abundances.
\item We have found some differences when applying exactly the same modeling to different atlases. Therefore, we find it may be convenient to infer solar abundances using the largest panoply of observations. However, we find a reliable oxygen abundance with just 0.003 dex difference between the studies with the two solar atlases.
\item A reliable independent determination of the spectral line atomic parameters is very important. Otherwise,
this uncertainty propagates accordingly onto the inferred abundances.
\item Given the interdependencies among all model parameters and the uncertainty
in the observations, one should pursue a fully Bayesian approach taking into account
all effects simultaneously. Ideally, such an approach should also be considered for aggregating
all sources of uncertainty and checking, through a meta-analysis, for the compatibility of all the inferred abundances
that exist in the literature. 
\end{itemize}
 
Despite the many factors that we take into account, we found a good agreement in our results for the 
oxygen abundance. Thus, we can conclude that, based on the three-dimensional model that we used, the most probable value for the oxygen solar abundance is $log(\epsilon_0) = 8.86\pm 0.04$. This value is classified as a high solar oxygen 
abundance and it is compatible with the results of \cite{1998SSRv...85..161G}.

Finally, a caveat is in order. Note that all results given in this paper are conditioned
on the empirical model of \cite{2011A&A...529A..37S,2015ascl.soft08002S} and the radiative
transfer of NICOLE being correct. If this is not the case, the conclusions might change. The
advantage of the Bayesian framework is that it is transparent (every probability distribution
in Eq. (\ref{eq:bayes1}) is conditioned on the a-priori information $I$).

\begin{acknowledgements}
Financial support by the Spanish Ministry of Economy and Competitiveness 
through projects AYA2014-60476-P and Consolider-Ingenio 2010 CSD2009-00038 
are gratefully acknowledged. AAR acknowledges financial support through the Ram\'on y Cajal fellowship. MCA acknowledges Fundaci\'on la Caixa for the financial support received in the form of a PhD contract. 
This research has made use of NASA's Astrophysics Data System Bibliographic Services. This work has made use of the VALD database, operated at Uppsala
University, the Institute of Astronomy RAS in Moscow, and the
University of Vienna.
We acknowledge the community effort devoted to the development of the following open-source Python packages that were
used in this work: \texttt{numpy}, \texttt{matplotlib}, \texttt{nestle}, and \texttt{seaborn}.
 \end{acknowledgements}

\nocite{2015PhyS...90e4005R}
\nocite{2000BaltA...9..590K}
\nocite{1999A&AS..138..119K}
\nocite{1997BaltA...6..244R}
\nocite{1995A&AS..112..525P}

\bibliographystyle{aa} 


\end{document}